\documentclass[12pt,twoside]{article}
\usepackage{fleqn,espcrc1,latexsym}

\usepackage{graphicx}

\newcommand{\AmS}{{\protect\the\textfont2
  A\kern-.1667em\lower.5ex\hbox{M}\kern-.125emS}}

\title{Bridging the mass gaps at $A=5$ and $A=8$ in
nucleosynthesis\thanks{This work was
partly supported by the Fonds zur wissenschaftlichen Forschung in \"Osterreich
(P13246-TPH), D32513/FKFP-0242-2000/BO-00520-98 (Hungary) and by the John
Templeton
Foundation (938-COS153). We would like to thank A. Weiss for 
useful discussions, and B. Gr\"un, S. Haslinger
and T. Strodl for their assistance in preparing this paper.}}
\author{H. Oberhummer\address{Institute of Nuclear Physics, Vienna University
of Technology,\\ Wiedner Hauptstra{\ss}e 8-10, A-1040 Vienna, Austria},
A. Cs\'ot\'o\address{Department of Atomic Physics, E\"otv\"os University, \\
P\'azm\'any P\'eter s\'et\'any 1/A, H-1117 Budapest, Hungary}, and
H. Schlattl\address{Max-Planck-Institut f\"ur Astrophysik,\\
Karl-Schwarzschild-Stra{\ss}e 1, D-85741 Garching, Germany}}

\begin{document}

\maketitle

\begin{abstract}
In nucleosynthesis three possible paths are known to bridge the mass gaps
at $A=5$
and $A=8$. The primary path producing the bulk of the carbon in our
Universe proceeds
via the triple-alpha process $^4$He(2$\alpha$,$\gamma$)$^{12}$C. This
process takes
place in helium-burning of red giant stars. We show that outside a narrow
window of
about 0.5\,\% of the strength or range of the strong force, the stellar
production of carbon or
oxygen through
the triple-alpha process is reduced by factors of 30 to 1000. Outside this
small
window the creation of carbon or oxygen and therefore also carbon-based
life in the
universe is strongly disfavored. The anthropically allowed strengths of the
strong
force also give severe constraints for the sum of the light quark masses as
well as
the Higgs vacuum
expectation value and mass parameter at the 1\,\% level.
\end{abstract}

\section{INTRODUCTION}

Few-body methods are used in nuclear astrophysics for
the determination of thermonuclear cross sections and reaction
rates predominantly for nuclei with mass numbers up to about $A=12$.
These reactions are of relevance for primordial nucleosynthesis, i.e.,
the production of nuclei up to $A=7$ in standard Big-Bang model.
In the inhomogeneous Big-Bang nucleosynthesis model
nuclei with $A > 7$ can also be produced.
Few-body methods play also a role in the determination of reaction
rates for the pp-chains in main-sequence stars up to $A=7$ and helium
burning from $A=4$ to $A=12$. Finally, in the alpha-rich freeze-out
occurring in supernovae, reaction rates for nuclei from $A=4$ to $A=12$
are studied using few-body methods.
\begin{figure}[tb]
\centering
\includegraphics[height=8cm]{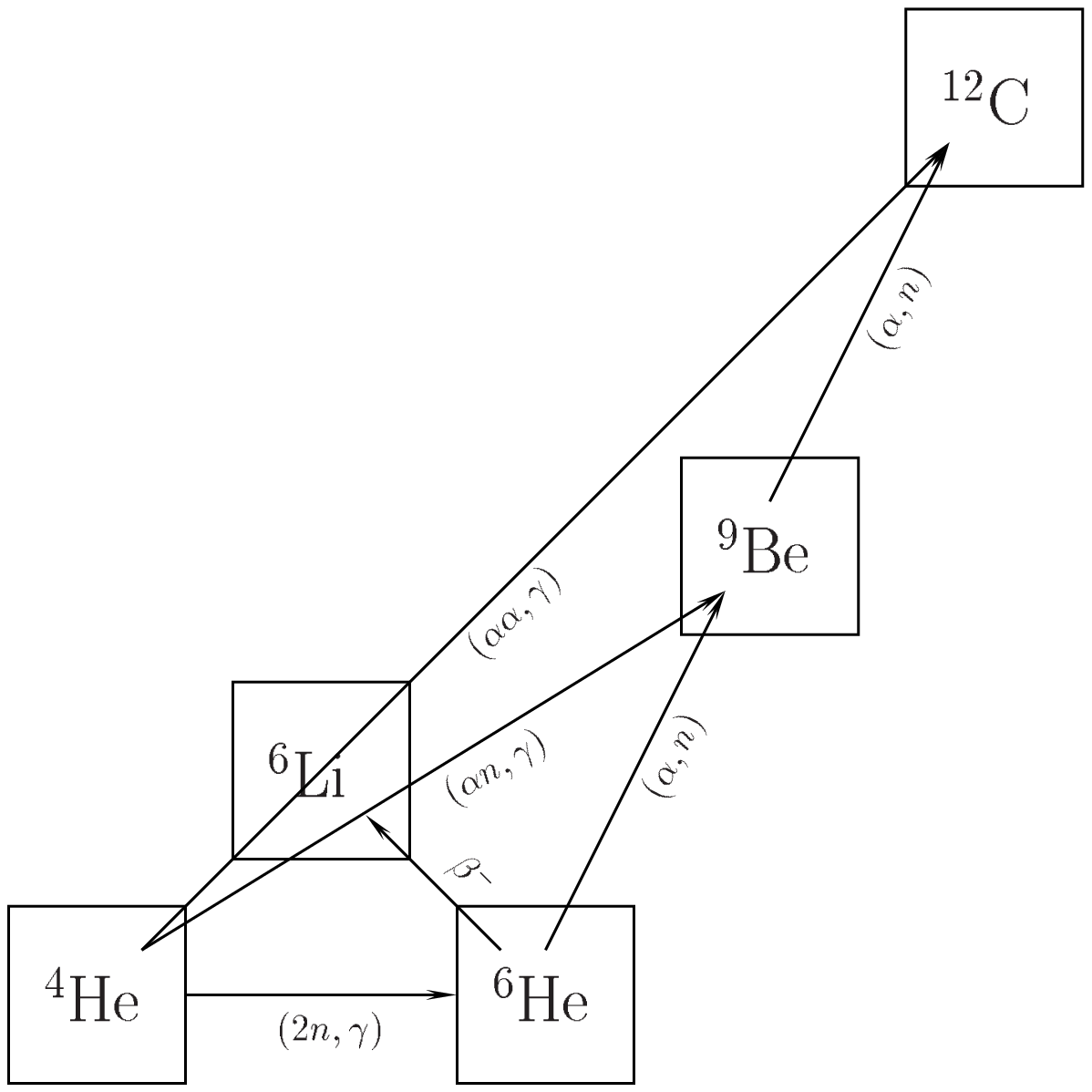}
\caption{Possible reaction paths bridging the mass numbers $A=5$ and $A=8$.}
\label{Fig1}
\end{figure}

We want to discuss the bridging of the mass gaps at $A=5$ and $A=8$ in
nucleosynthesis through three-body reactions (Fig.~\ref{Fig1}).
The following paths via three-body reactions can bridge these mass gaps:
\begin{enumerate}
\item Reaction path
$^4$He(2n,$\gamma$)$^6$He($\alpha$,n)$^9$Be($\alpha$,n)$^{12}$C:
Under representative conditions
the reaction path
$^4$He(2n,$\gamma$)$^6$He($\alpha$,n)$^9$Be($\alpha$,n)$^{12}$C
can be neglected in known astrophysical scenarios, because the fragile $^6$He
is destroyed very effectively by photodissociation
$^6$He($\gamma$,2n)$^4$He \cite{gor95,efr96,her99}.
The cross sections and reaction rates for the reaction
$^4$He(2n,$\gamma$)$^6$He
and the reverse photodissociation have been calculated
using an $\alpha$+n+n approach \cite{efr96}.
\item Reaction path $^4$He($\alpha$n,$\gamma$)$^9$Be($\alpha$,n)$^{12}$C:
This reaction path is dominant in the
so-called alpha-rich freeze-out occurring in type II supernovae
\cite{gor95,woo94,tak94,woo92,efr98}.
The cross sections and the reaction rates for $^4$He($\alpha$n,$\gamma$)$^9$Be
and the reverse photodissociation have been
calculated in a semimicroscopic model and compared with the
experimental data \cite{efr98}.
\item Triple-alpha process $^4$He(2$\alpha$,$\gamma$)$^{12}$C:
The bulk of the carbon in the universe is produced through this process
in the helium-burning phase of red giant stars.
The reaction rate for the triple-alpha process
has been determined recently in a microscopic 12-nucleon model
\cite{obe00a,obe00b,cso00}.
\end{enumerate}

In the following section we will present the calculation
of the reaction rate for the triple-alpha process in the microscopic
12-nucleon model
$^4$He(2$\alpha$,$\gamma$)$^{12}$C.
Sect.~3 will be devoted to variations
of the nucleon-nucleon (N-N) force and its consequences for the stellar
production of carbon and
oxygen in the universe. In Sect.~4 we discuss
the constraints for anthropically allowed
parameters of elementary particle physics like the light quark masses
or the Higgs vacuum expectation value and mass parameter.
In the last section a short summary
is given.

\section{REACTION RATE FOR THE TRIPLE-ALPHA PROCESS}

The formation of $^{12}$C in hydrogen burning
is blocked by the absence of stable elements at
mass numbers $A=5$ and $A=8$. \"Opik \cite{OP51} and Salpeter \cite{SA52}
pointed out
that the lifetime of $^8$Be is long enough,
so that the $\alpha +\alpha \rightleftharpoons\; $$^8$Be
reaction can produce macroscopic amounts of equilibrium
$^8$Be in stars. Then, the unstable $^8$Be could capture an
additional $\alpha$ particle to produce stable $^{12}$C.
However, this so-called triple-alpha reaction has very
low rate since the density of $^8$Be in the stellar
plasma is very low, because of its short lifetime of 10$^{-16}$\,s.

Hoyle (cited in \cite{DU53}) argued that the triple-alpha
reaction cannot produce enough carbon in a non-resonant way
in order to explain the measured abundance in the Universe, therefore
it must proceed through a hypothetical resonance of
$^{12}$C, thus strongly enhancing the cross section.
Hoyle suggested that this resonance is a $J^\pi=0^+$ state
at about $\varepsilon=0.4$\,MeV (throughout this paper $\varepsilon$ denotes
resonance energy in the center-of-mass frame relative to
the three-alpha threshold, while $\Gamma$ denotes the full
width). Subsequent experiments indeed found a $0^+$
resonance in $^{12}$C in the predicted energy region by
studies of the reaction $^{14}$N(d,$\alpha$)$^{12}$C \cite{DU53,HO53}
and the $\beta^-$-decay of $^{12}$B \cite{CO57}. It
is the second $0^+$ state ($0^+_2$) in $^{12}$C. Its modern
parameters, $\varepsilon=0.3796$\,MeV and
$\Gamma=8.5 \times 10^{-6}$\,MeV ~\cite{Ajzenberg}, agree well with the old
theoretical prediction.

\subsection{Microscopic 12-body model}

The astrophysical models that determine the amount of carbon and oxygen
produced in red giant stars need some nuclear properties of $^{12}$C as
input parameters. Namely, the position and width of the $0^+_2$
resonance, which almost solely determines the triple-alpha reaction
rate, and the radiative decay width for the $0^+_2\rightarrow 2^+_1$
transition in $^{12}$C. We calculated these quantities in a
microscopic 12-body model \cite{obe00a,obe00b,cso00}.

In the microscopic cluster model it is assumed that the wave functions of
certain
nuclei, like $^{12}$C, contain, with large weight, components which
describe the given nucleus as a compound of 2-3 clusters. By assuming
rather simple structures for the cluster internal states, the relative
motions between the clusters, which are the most important degrees of
freedom, can be treated rigorously. The strong binding of the free
alpha particle $^4$He makes it natural that the low-lying states of
$^{12}$C have $3\alpha$-structures \cite{3alpha}. Therefore, our
cluster-model wave function for $^{12}$C looks like
\begin{equation}
\label{wfn}
\Psi^{^{12}{\rm C}}=\sum_{l_1,l_2} {\cal A} \Bigl
\{ \Phi^\alpha \Phi^\alpha \Phi^\alpha\chi^{\alpha(
\alpha\alpha)}_{[l_1l_2]L} (\mbox{\boldmath$\rho$}_1,
\mbox{\boldmath$\rho$}_2) \Bigl\}.
\end{equation}
Here ${\cal A}$ is the intercluster antisymmetrizer, the $\Phi^\alpha$
cluster internal states are translationally invariant $0s$
harmonic-oscillator shell-model states with zero total spin, the
$\mbox{\boldmath$\rho$}$ vectors are the intercluster Jacobi
coordinates, $l_1$ and $l_2$ are the angular momenta of the two
relative motions, $L$ is the total orbital angular momentum and
$[\ldots]$ denotes angular momentum coupling. The total spin and
parity of $^{12}$C are $J=L$ and $\pi=(-1)^{l_1+l_2}$, respectively.

We performed calculations using the Minnesota (MN)
and modified Hasegawa-Nagagta (MHN) N-N forces.
These two forces give the best simultaneous description of the $^8$Be 
ground state and the $0^+_2$ state of $^{12}$C \cite{cso00}, in 
agreement with the experience gained
in the cluster-model description of the structure and reactions of various
light nuclei \cite{lan94}.
The MN-force is given by \cite{MN1,MN2}
\begin{equation}
\label{MN}
V_{ij}(r)=\left(V_1(r)+{{1}\over{2}}(1+P_{ij}^\sigma)
V_2(r)+{{1}\over{2}}(1-P_{ij}^\sigma)V_3(r) \right)
\left({{1}\over{2}}u+{{1}\over{2}}(2-u)P_{ij}^r \right)+V_{ij}^{\rm Coul.}(r),
\end{equation}
where $P^r$ and $P^\sigma$ are the space- and spin-exchange operators,
respectively, $u$ is the exchange mixture parameter, $r=\vert {\bf r}_j-{\bf
r}_i\vert$, and $V_{ij}^{\rm Coul.}$ is the Coulomb force between the two
nucleons.
The MHN force is given by \cite{MHN}
\begin{equation}
\label{MHN}
V_{ij}(r)=\sum_{k=1}^3\Big (W_k+M_kP^r_{ij}+B_kP^{\sigma}_{ij}
+H_kP^{\tau}_{ij}\Big )V_k(r)+V_{ij}^{\rm Coul.}(r).
\end{equation}
The parameter $W_k$ is the Wigner parameter,
$P^r_{ij}$, $P^{\sigma}_{ij}$, and $P^{\tau}_{ij}$ are the space-, 
spin-, and isospin-exchange operators (Majorana-, Bartlett-, and
Heisenberg-operators), and $M_k$, $B_k$, and $H_k$ are the corresponding
exchange mixture parameters.

The spatial parts of the MN- and MHN-force have Gaussian forms
\begin{equation}
\label{NNP}
V_k(r)=V_{0k}\exp \left[- \left(\frac{r}{r_{0k}}\right)^2\right],\ \  k=1,2,3,
\end{equation}
where $V_{0k}$ and $r_{0k}$ are the strength and range parameters
of the potentials, respectively.

\subsection{Triple-alpha reaction rate}

The resonant reaction rate for the triple-alpha process proceeding via the
ground state of $^8$Be and the $0^+_2$ resonance in $^{12}$C is given
approximately by \cite{rol88}
\begin{equation}
\label{alphaa}
r_{3\alpha} \approx 3^{\frac{3}{2}} N_{\alpha}^3
\left(\frac{2 \pi \hbar^2}{M_{\alpha} k_{\rm B} T}\right)^3
\frac{\Gamma_{\gamma}}{\hbar} \exp \left(- \frac{\varepsilon}{k_{\rm B} T}
\right),
\end{equation}
where $M_{\alpha}$ and $N_{\alpha}$ are the mass and the number density of
the alpha particle, while $\hbar$ and $k$ are Planck's and Boltzmann's
constant, respectively. The temperature of the stellar plasma
is given by $T$.

We calculated the resonance energy $\varepsilon$ of the $0^+_2$ state in
$^{12}$C, relative to the $3\alpha$-threshold, and the
$0^+_2\rightarrow 2^+_1$ radiative (E2) width $\Gamma_{\gamma}$. The
$0^+_2$ state
is situated above the
$3\alpha$-threshold, therefore for a rigorous description one has to
use an approach which can describe three-body resonances correctly. We
choose the complex scaling method \cite{CSM} that has already been used
in a variety of other nuclear physics problems, see
e.g.~\cite{3alpha,CSMa,CSMb}.

As a first step of our calculations,
we fine tune each N-N force (by slightly
changing their exchange-mixture parameters) to fix $\varepsilon$ in
Eq.~(\ref{alphaa}) at its experimental value. The other important quantity that
needs to be calculated is the
radiative width of the $0^+_2$ state, coming from the electric dipole
(E2) decay into the $2^+_1$ state of $^{12}$C. This calculation involves
the evaluation of the E2 operator between the initial $0^+_2$
three-body scattering state and the final $2^+_1$ bound state \cite{Gamma}.
The proper three-body scattering-state treatment of the $0^+_2$
initial state is not feasible in our approach for the time being, 
therefore we use a
bound-state approximation to it. This is an excellent approximation for
the calculation of $\Gamma_\gamma$ because the total width of the
$0^+_2$ state is very small (8.5\,eV \cite{Ajzenberg}). The value of
$\Gamma_\gamma$ is rather sensitive to the energy difference between the
$0^+_2$ and $2^+_1$ states, so we have to make sure that the experimental
energy difference is correctly reproduced.

\subsection{Stellar-model calculations}

The composition of the interstellar material (ISM) is a mixture of
ejecta from stars with different masses. At present it is not clear
which type of stars contribute most of the $^{12}$C or $^{16}$O to the
ISM. Therefore, we performed stellar model calculations for a typical
massive, in\-ter\-me\-di\-a\-te-mass and low-mass star with masses 
20, 5, and 1.3\,$M_\odot$, respectively, including the calculated
triple-alpha reaction rates. 

We used a modern stellar evolution code, which contains the most
recent input physics \cite{WS2000}. Up-to-date solar models can be
calculated with this program \cite{SW99} as well as the evolution of
low-mass stars can be followed through the thermal-pulse phase of
stars at the asymptotic giant branch \cite{WaWe94}. The nuclear
network is designed preferentially to calculate the H- and
He-burning phases in low-mass stars. Additionally, all
basic reactions of C- and O-burning are included, which may destroy
the previously produced C and O in massive and intermediate-mass stars.

Here, the stars are followed from the onset of H-burning until the third
thermal pulse on the AGB, or until the core temperature reaches $10^9$\,K in
the case of the 20\,$M_{\odot}$ star (the nuclear network is not sufficient
to go beyond this phase). 

Large portions of the initial mass of a star are returned to the ISM
through stellar winds. Unfortunately, basically due to the simple
convection model used in stellar modeling, the composition of the wind
cannot yet be determined very accurately from stellar evolution
theory.  However, it is beyond the scope of the present investigations to
determine how and when the material is returned to the ISM.
Instead we examine how much C and O is produced altogether.

For the 1.3\,$M_\odot$ star, which loses its envelope basically during
the thermal-pulse phase, the maximum C and O abundances in the
He-burning region have been extracted. Although the efficiency of the 
dredge-up of heavy elements to the surface is only badly known, it is
independent of the nuclear physics, and hence should be similar in all
models independent of the triple-alpha rate.
By taking the maximum abundances in this region, we have a measure of 
how strongly the enrichment of the stellar envelope by C or O is 
altered by modifying the triple-alpha rate.

In the 5 and 20\,$M_\odot$ stars further fusion reactions like C-
and O-burning take place. The dredge-up process of metal-enriched
material under these circumstances is more complicated and more
uncertain than in the 1.3\,$M_\odot$ star. The 20\,$M_\odot$ star
finally even explodes in a supernova.
Therefore, for the 5 and 20\,$M_\odot$ stars the total amount of C and O in the stellar interior is evaluated.

\section{VARIATIONS OF THE NUCLEON-NUCLEON FORCE}

In this section we slightly vary the strength and range
parameters of the MN- and MHN-potentials
and calculate the modified resonance energies $\varepsilon$
and gamma widths $\Gamma_{\gamma}$ of the $0^+_2$ state in $^{12}$C.
With these values of $\varepsilon$ and $\Gamma_{\gamma}$
we recalculate the triple-alpha reaction rate $r_{3\alpha}$
of Eq.~(\ref{alphaa}). We find that the value of $\Gamma_\gamma$ 
is very little changed by the small
variations of the N-N interactions, leading to negligible changes in
$r_{3\alpha}$. Thus, in the stellar model calculations we can fix
$\Gamma_\gamma$ to its experimental value in all cases. The resonance energy
$\varepsilon$, however, is rather sensitive to variations in the N-N force,
leading to large changes in the triple-alpha rate $r_{3\alpha}$. By
making the N-N force weaker, the resonance energy $\varepsilon$ moves
higher and the reaction rate $r_{3\alpha}$ decreases exponentially as can
be seen from Eq.~(\ref{alphaa}). The opposite is true for a stronger N-N force.

We have done a series of tests in order to make sure that the main
assumptions of our calculations remain valid while the interactions are
varied. Here we mention only three main points.
We checked that the use of Eq.~(\ref{alphaa}) is justified 
for all the different resonance energies as well as
temperatures occurring in  our calculations. In all cases the triple-alpha
reaction is dominated by the sequential process through the $^8$Be ground
state and the $0^+_2$ state of $^{12}$C. The amount of variations in the
strength of the Coulomb and strong interaction are
small enough to keep the experimental $^8$Be ground state and the $0^+_2$
state of $^{12}$C from becoming a broad state or a bound state. We also 
estimated the non-resonant contribution \cite{lan86}, without the $0^+_2$
resonance in $^{12}$C, to the triple-alpha rate and found that this
contribution is about 7-13 orders of magnitude smaller than the resonant
contribution in the stellar temperature range $T \approx (0.8-3) \times
10^8$\,K arising in our stellar-model calculations. Furthermore, the
contribution of the next higher-energy $0^+$ resonance (at about 2.7\,MeV
above the three-alpha threshold \cite{rol88}) to the triple-alpha rate is
smaller by
about 40 orders of magnitude in the considered temperature range. Also, in the
above temperature range the reaction will not proceed through the low-energy
wing of the $^8$Be resonance \cite{lan86}.

Some of the carbon, produced in the triple-alpha process, is further
synthesized in the $^{12}{\rm C}(\alpha,\gamma){^{16}{\rm O}}$ and
$^{16}{\rm O}(\alpha,\gamma){^{20}{\rm Ne}}$ reactions.
The $^{16}{\rm O}(\alpha,\gamma){^{20}{\rm Ne}}$ reaction is nonresonant, so
variations in the strengths of the strong force can have only small
effects on its reaction rate. The $^{12}{\rm C}(\alpha,\gamma){^{16}{\rm O}}$
process may look more dangerous because its cross section is strongly
affected by subthreshold states in the $^{16}$O nucleus \cite{rol88}.
However, if the
N-N force is made weaker, then the subthreshold states become less bound,
thereby enhancing the $^{12}{\rm C}(\alpha,\gamma){^{16}{\rm O}}$ cross
section.
Therefore, in the case of a weaker force the small C/O ratio is further
decreased. An analogous reasoning holds for a stronger force. Thus, without
doing any calculation for the  $^{12}{\rm C}(\alpha,\gamma){^{16}{\rm
O}}$ and $^{16}{\rm O}(\alpha,\gamma){^{20}{\rm Ne}}$ reactions with the
modified forces, we can conclude that their effect would strengthen our
hypothesis regarding carbon and oxygen production.

There may be other windows for different values of the fundamental 
interactions in which an appreciable amount of carbon and oxygen could be 
produced. The window that seems to be closest to the values realized in 
our universe is the creation of oxygen and carbon in the big 
bang \cite{alp48}. This possibility is prevented in our universe
by the existence of the stability gaps at mass numbers $A=5$ and $A=8$.
In order to make $^5$He or $^5$Li as well as $^8$Be become bound, we
calculated that the strong forces would have to be
increased by about 10\%, a value that is about a factor 20 larger than
the maximal variation investigated in this work.

We multiply now in the MN- and MHN-potentials independently
the strength parameters $V_{0k}$
as well as the range parameters $r_{0k}$
for all repulsive and attractive terms $V_{0k}$
in Eq.~(\ref{NNP}) by the same factor
$p$. This factor was set between 0.994 and 1.006 for the
strength parameters and 0.997 and 1.003 for the
range parameters of the MN and MHN potentials. With
these values for the factor $p$ we
recalculate the resonance energies and the triple-alpha reaction rates.
For each of these new reaction rates we then perform again
the corresponding stellar model calculations.

\begin{figure}[tb]
\centering
\includegraphics[width=14cm]{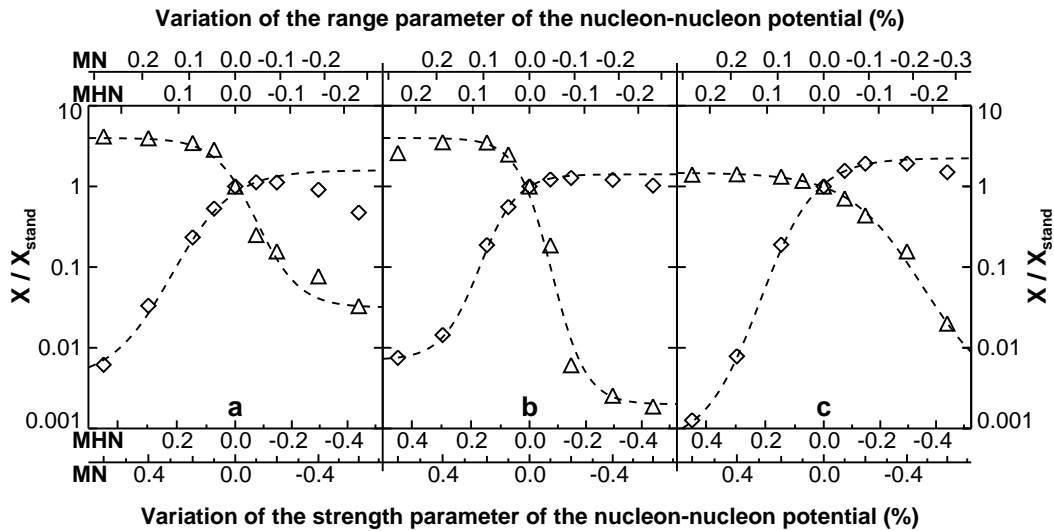}
\caption{The change of the carbon ($\triangle$) and oxygen
($\Diamond$) mass abundances ($X$) through variations of the strength
(lower ordinate) and range parameters (upper ordinate) of the
nucleon-nucleon potentials MN and MHN.
The abundance changes are shown in panels a, b, and c for
stars with masses of 20, 5, and 1.3\,$M_\odot$, respectively, in units
of the standard values $X_{\rm stand}$. The dashed curves are drawn to
guide the eye.}
\label{Fig2}
\end{figure}

The resulting changes in the C and O abundances are shown
(Fig.~\ref{Fig2}) with respect to the case, where the standard value of
the resonance energy has been used (i.e., with no variations of the 
strength or range parameters of the MN- and MHN-potentials).
Because each shift in the resonance energy can be
identified with a variation in the strength or range parameter of the
corresponding N-N force,
we scaled the lower and upper ordinate with variations in these quantities,
respectively.

We conclude that a change of more than about 0.5\,\% in the strength
parameters or
0.2-0.3\,\%
in the range parameters of the MHN- or MN-potential would destroy either
nearly all C or
all O in every star. This implies that irrespective of
stellar evolution the contribution of each star to the abundance
of C or O in the ISM would be negligible. This corresponds to a fine tuning
that is
about two orders of magnitude better than obtained from the constraints of
a bound deuteron
\cite{poc71,agr98a,agr98b} or from the non-existence of a bound diproton or
dineutron \cite{dys71}.

In Fig.~\ref{Fig2} one can see that the size of the resulting change in the
abundances with the variation of the range parameters
$r_{0k}$ is about twice the size of the change coming from the variation of
the strength parameters $V_{0k}$.
This can be understood quite simply, because we only
considered small variations ($\approx\,$0.5\,\%) of
the strength and range parameters, respectively. In this case
the changes in the strength and range parameters in Eq.~(\ref{NNP})
\begin{eqnarray}
\label{Var1}
V_{0k}^{\prime} & = & (1 \pm \delta V_{0k}) V_{0k},
\nonumber \\
r_{0k}^{\prime} & = & (1 \pm \delta r_{0k}) r_{0k}
\end{eqnarray}
lead to a change in the spatial parts in the potential
for 
$\delta r_{0k} \ll 1$
\begin{eqnarray}
\label{Var2}
V_{k}^{\prime}(r) &
= & V_{0k}^{\prime} \exp \left[- \left(\frac{r}{r_{0k}}\right)^2\right]
= (1 \pm \delta V_{0k}) V_{k}(r), \nonumber \\
V_{k}^{\prime}(r) &
= & V_{0k} \exp \left[- \left(\frac{r}{r_{0k}^{\prime}}\right)^2\right]
\approx V_{0k} \exp \left[- (1 \mp 2 \delta r_{0k})
\left(\frac{r}{r_{0k}}\right)^2\right]
\approx (1 \pm 2 \delta r_{0k}) V_{k}(r),
\end{eqnarray}
where in the second line we used a mean potential range parameter
$r \approx r_{0k}$ in the first-order correction term. Therefore, in the
above approximation a change in the range parameter $r_{0k}$ corresponds
to a twice as large change, $\delta V_{0k} \approx + 2 \delta r_{0k}$,
in the strength parameter.

\section{ANTRHOPICALLY ALLOWED QUARK MASSES AND HIGGS PA\-RA\-ME\-TERS}

In this section we investigate the constraints on the anthropically allowed
quark masses
following largely the discussion already given in Ref.~\cite{hog99}.

The one-boson exchange potential (OBEP) for the N-N force can be written as
a sum of Yukawa terms
\begin{equation}
\label{OBEP}
V_{\rm OBEP}(r) =
- f_{\pi}^2\frac{\exp(- m_{\pi} r)}{r} -
f_{\sigma}^2 \frac{\exp(- m_{\sigma} r)}{r} +
f_{\omega}^2 \frac{\exp(- m_{\omega} r)}{r},
\end{equation}
where the f's and m's are the coupling constants and masses of the $\pi$,
$\sigma$
and $\omega$ mesons, respectively.
The OBEP-potential has a long-range tail from one-pion exchange, an
attractive minimum
due to the exchange of the hypothetical $\sigma$-meson and a repulsive core
due to
the $\omega$-meson. In the following we will only consider the long-range
part due to
the pion, the one-pion exchange potential (OPEP),
\begin{equation}
\label{OPEP}
V_{\rm OPEP}(r) = - f_{\pi}^2 \frac{\exp(- m_{\pi} r)}{r}.
\end{equation}

For small changes of the pion mass,
\begin{equation}
\label{Var3}
m_{\pi}^{\prime} = m_{\pi} \pm \delta m_{\pi}, 
\end{equation}
we obtain
\begin{equation}
\label{Var4}
V_{\rm OPEP}^{\prime}(r) =
-f_{\pi}^2 \frac{\exp(- m_{\pi}^{\prime} r)}{r} \approx
-f_{\pi}^2 \frac{\exp[- (m_{\pi} \pm \delta m_{\pi})r)]}{r}
\approx \left(1 \mp \frac{\delta m_{\pi}}{m_{\pi}}\right) V_{\rm OPEP}(r),
\end{equation}
where in the second line we used a mean potential range given
by the pion mass $r \approx 1/m_{\pi}$ in the first-order
correction term.

The range defined by the inverse pion-mass is therefore also limited
to a $\pm0.5\%$ window. Here we assumed that the sensitivity of the
carbon production to the potential strength is similar in the
OPEP and MN/MHN cases, respectively.

As long as the up- and down-quark masses are small compared to the
QCD scale, the mass of the pion is well approximated by
$m_{\pi} \propto \sqrt{f_{\pi}(m_{\rm u}+m_{\rm d})}$ \cite{agr98a,agr98b}.
Therefore, the  sum of the up- and down-quark masses
$m_{\rm u+d} \equiv m_{\rm u} + m_{\rm d}$
scales with the pion mass $ m_{\pi}$ as $0.5 \delta m_{\rm u+d} \approx
\delta m_{\pi}$.
Thus the anthropically allowed value of the sum of the up- and down-quark
masses are fine tuned to approximately 1\,\%.
Using the quark masses $m_{\rm u} =  (4.88 \pm 0.57)$\,MeV and
$m_{\rm d} = (9.81 \pm 0.65)$\,MeV the anthropically allowed
sum of the up- and down-quark masses $m_{\rm u+d}$ is constrained to
within about 0.15\,MeV. This is consistent with the result of
Ref.~\cite{hog99}
where a value in the order of 0.05\,MeV is given.

Constraints can also be obtained for the difference of the down- and
up-quark masses, $m_{\rm d-u}\equiv m_{\rm d}-m_{\rm u}$. 
We do not consider possible variations of the electron mass
as in Ref.~\cite{hog99}. The constraint for
the upper limit of $\delta m_{\rm d-u}$ follows
from the restriction that the pp-fusion
\begin{equation}
\label{pp}
{\rm p} + {\rm p} \rightarrow {\rm d} + {\rm e}^+ + \nu
\end{equation}
is exothermic. This gives an allowed range of $\delta m_{\rm d-u} < 0.42$\,MeV.
The constraint for
the lower limit of $\delta m_{\rm d-u}$ follows from the
stability of the hydrogen atom, i.e. that the proton capture of an electron
\begin{equation}
\label{pe}
{\rm p} + {\rm e}^- \rightarrow {\rm n} + \nu
\end{equation}
is endothermic. This gives an allowed range of $\delta m_{\rm d-u} >
-0.782$\,MeV.
The stability of the proton and deuteron give also constraints for the lower
and upper limit of $\delta m_{\rm d-u}$. However, these constraints
are less stringent than the ones obtained from the ones given above
\cite{hog99}.

\begin{figure}[tb]
\centering
\includegraphics[width=9.0cm]{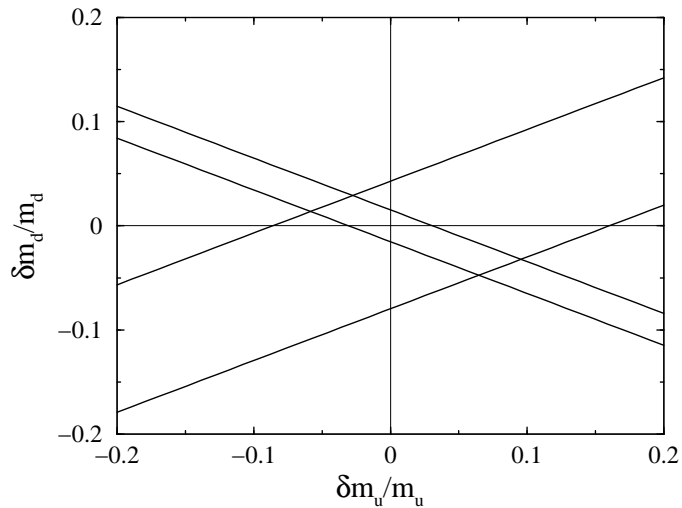}
\caption{The central region shows the anthropically allowed values
of variations of the up-quark mass $\delta m_{\rm u}/m_{\rm u}$ and
variations of the down-quark mass $\delta m_{\rm d}/m_{\rm d}$.
The following constraints are determined by the straight lines
in the figure: (i) the pp-fusion in Eq.~(\ref{pp}) is exothermic,
(ii) the proton capture of an electron in Eq.~(\ref{pe}) is
endothermic, (iii) the carbon production is not suppressed, and (iv)
the oxygen is production is not suppressed. These four constraints
determine the upper and lower bounds of the bands around
$\delta m_{\rm d}/m_{\rm d} = + 0.5 \delta m_{\rm u}/m_{\rm u}$ and
$\delta m_{\rm d}/m_{\rm d} = - 0.5 \delta m_{\rm u}/m_{\rm u}$,
respectively.}
\label{Fig3}
\end{figure}

The resulting anthropically allowed strengths of the up- and
down-quark masses are given by the central region of Fig.~\ref{Fig3}.
As can be seen from this figure the anthropic fine tuning
of the up- and down-quark masses is less than a few percent.
The band around $\delta m_{\rm d}/m_{\rm d} =
-0.5\delta m_{\rm u}/m_{\rm d}$
determined by the production of carbon and oxygen is about four times  
narrower than the band around
$\delta m_{\rm d}/m_{\rm d} = + 0.5 
\delta m_{\rm u}/m_{\rm u}$. 
That means that the constraints of carbon or oxygen production give 
a stronger
fine tuning than the constraints determined from the reactions
of Eqs.~(\ref{pp}) and (\ref{pe}). If additional joint constraints
that unification imposes on the fermion masses like the ratio $m_{\rm
d}/m_{\rm e}$
are taken into account \cite{fug99}, three-dimensional
constraining plots for
the up-quark, down-quark and electron masses can be derived.

Similar considerations \cite{agr98a,agr98b,jel99} can be carried out for
the Higgs
vacuum expectation value $v \approx \sqrt{-\mu^2/\lambda} \propto \mu$
and the Higgs mass parameter $\mu$. Since the up- and down-quark
masses scale simply with the Higgs vacuum expectation value,
the Higgs vacuum expectation value $v$
and Higgs mass parameter $\mu$ are also fine tuned by anthropic considerations
to approximately $\pm 1$\,\%.

\section{SUMMARY}

In this work we have shown that the strength or range of the strong force is
fine tuned to $\pm 0.5$\,\% by the production of carbon or oxygen and
therefore also carbon-based life. This leads to a fine tuning of the
sum of the light quark masses as well as the Higgs vacuum expectation value
and mass parameter to about $\pm 1$\,\%.

One of the most fascinating aspects
resulting from this is that life, stars, nuclei and elementary particles seem
to be closely interwoven in our universe through this extreme fine tuning.
It is not clear if the values of the fundamental parameters corresponding 
to this
anthropic fine tuning will be calculable by a future Final Theory \cite{per00}
or if some of the parameters of such a Final Theory
will have to be chosen from a large or continuous ensemble \cite{hog99}. In
any case it will be one of the challenges for a possible Final Theory to
explain the
anthropic fine tuning as described in this paper.

\end{document}